%% file: main.tex
\documentclass[aps,prx,superscriptaddress,twocolumn,showkeys]{revtex4-2}

\usepackage{graphicx}
\usepackage[font={small}]{caption}
\usepackage{subcaption}
\usepackage{multirow}
\usepackage[dvipsnames]{xcolor}
\usepackage{colortbl}
\usepackage{mathtools}
\usepackage{adjustbox}
\usepackage{amsmath}
\usepackage{listings}
\usepackage{multirow}
\usepackage{float}
\usepackage{url}
\usepackage{placeins} 

\begin{document}

\setlength{\tabcolsep}{4pt} 
\renewcommand{\arraystretch}{1.05}
\renewcommand{\thetable}{\arabic{table}}

\title{The role of electron correlations in the electronic structure of putative Chern magnet TbMn$_6$Sn$_6$}

\author{Abdulgani Annaberdiyev}
\affiliation{Center for Nanophase Materials Sciences, Oak Ridge National Laboratory, Oak Ridge, Tennessee 37831, USA}
\email{annaberdiyea@ornl.gov}

\author{Subhasish Mandal}
\affiliation{Department of Physics and Astronomy, West Virginia University, Morgantown, WV 26506, USA}
\email{subhasish.mandal@mail.wvu.edu}

\author{Lubos Mitas}
\affiliation{Department of Physics, North Carolina State University, Raleigh, North Carolina 27695-8202, USA}

\author{Jaron T. Krogel}
\affiliation{Materials Science and Technology Division, Oak Ridge National Laboratory, Oak Ridge, Tennessee 37831, USA}

\author{Panchapakesan Ganesh}
\affiliation{Center for Nanophase Materials Sciences, Oak Ridge National Laboratory, Oak Ridge, Tennessee 37831, USA }
\email{ganeshp@ornl.gov}

\begin{abstract}
A member of the RMn$_6$Sn$_6$ rare-earth family materials, TbMn$_6$Sn$_6$, recently showed experimental signatures of the realization of a quantum-limit Chern magnet.
In this work, we use quantum Monte Carlo (QMC) and density functional theory with Hubbard $U$ (DFT$+U$) calculations to examine the electronic structure of TbMn$_6$Sn$_6$.
To do so, we optimize accurate, correlation-consistent pseudopotentials for Tb and Sn using coupled-cluster and configuration-interaction (CI) methods.
We find that DFT$+U$ and single-reference QMC calculations suffer from the same overestimation of the magnetic moments as meta-GGA and hybrid density functional approximations.
Our findings point to the need for improved orbitals/wavefunctions for this class of materials, such as natural orbitals from CI, or for the inclusion of multi-reference effects that capture the static correlations for an accurate prediction of magnetic properties.
DFT$+U$ with Mn magnetic moments adjusted to experiment predict the Dirac crossing in bulk to be close to the Fermi level, within $\sim 120$~meV, in agreement with the experiments.
Our non-stoichiometric slab calculations show that the Dirac crossing approaches even closer to the Fermi level, suggesting the possible realization of Chern magnetism in this limit.
\end{abstract}


\maketitle

\section{Introduction}

\begin{figure}[htb]
\centering
\includegraphics[width=0.5\textwidth]{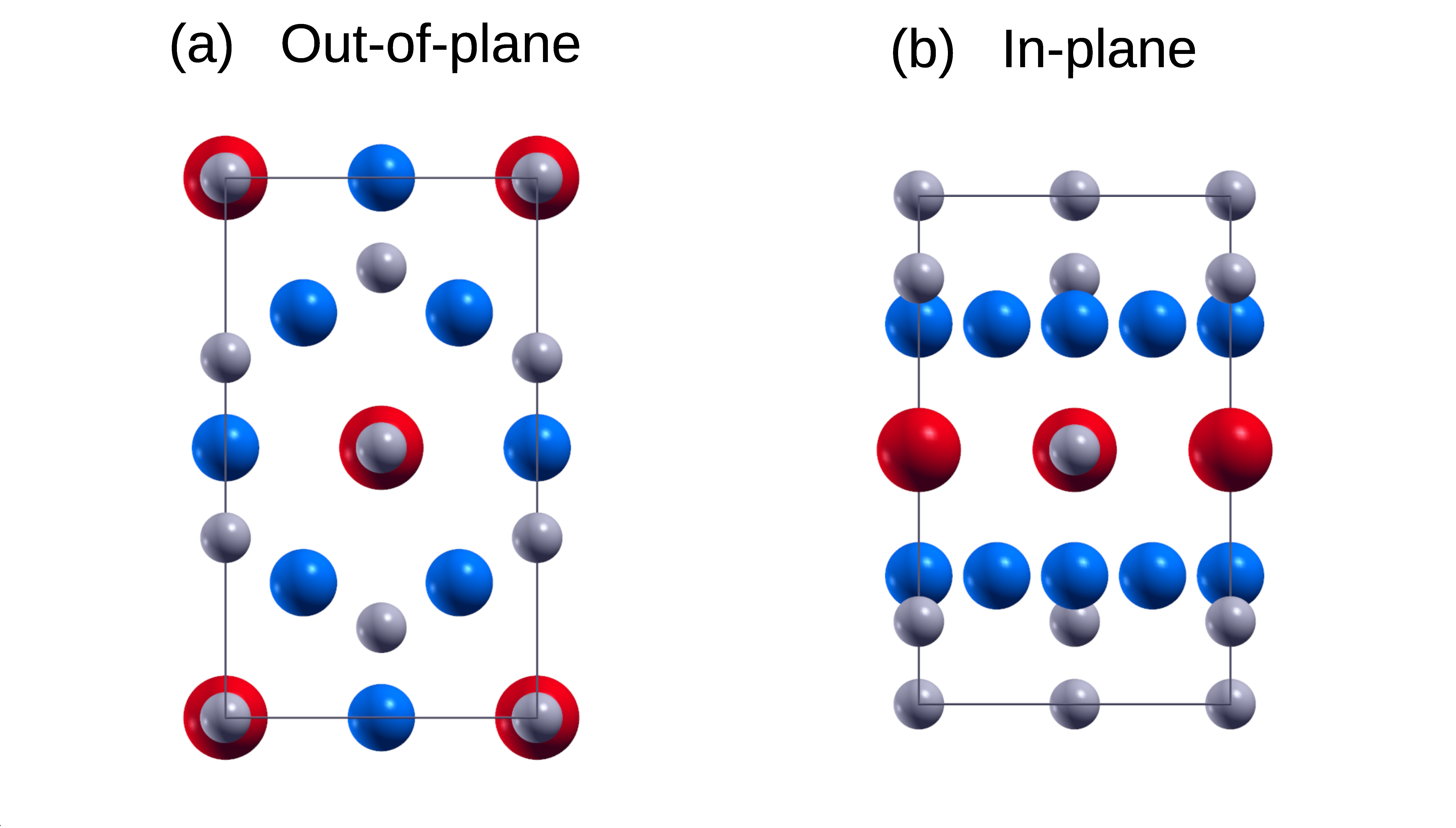}
\caption{
A supercell of TbMn$_6$Sn$_6$.
(a) Out-of-plane and (b) in-plane views are shown.
The large red spheres are Tb atoms, medium blue spheres are Mn atoms, and small gray spheres are Sn atoms.
\label{fig:qmc_cell}
}
\end{figure}

Rare-earth (R) material class RMn$_6$Sn$_6$ displays rich and intricate physical phenomena consisting of strong electron-electron correlations, spin-orbit effects, and supposedly topological behavior \cite{yin_quantum-limit_2020, lee_interplay_2022, xu_topological_2022, sims_evolution_2022, riberolles_low-temperature_2022, mielke_iii_low-temperature_2022, jones_origin_2022, ma_rare_2021, gao_anomalous_2021, ghimire_competing_2020}.
These materials belong to the $P6/mmm$ (\#191) space group, where Mn atoms form 2D kagome layers interlaced with R and Sn layers, see Figure \ref{fig:qmc_cell}.
Although RMn$_6$Sn$_6$ family of materials has been studied for more than three decades, recent experimental anomalous Hall effect (AHE) studies \cite{yin_quantum-limit_2020, gao_anomalous_2021, ma_anomalous_2021} inspired another wave of research \cite{ma_rare_2021, peng_realizing_2021, hu_tunable_2022, jones_origin_2022, lee_interplay_2022, li_selective_2022, mielke_iii_low-temperature_2022, min_topological_2022, riberolles_low-temperature_2022, sims_evolution_2022, xu_topological_2022, chen_large_2021, zhou_quantum_2022} on related materials.

Experimental studies show that stable compounds form for R = \{Sc, Y, Gd, Tb, Dy, Ho, Er, Tm, Yb, Ly\} \cite{kimura_high-field_2006, malaman_magnetic_1999, clatterbuck_magnetic_1999, venturini_incommensurate_1996, amako_the119sn_1994, venturini_magnetic_1991, el_idrissi_magnetic_1991, chafik_el_idrissi_refinement_1991}.
In this study, R = Tb is of particular interest since it is the only element of the previous set that forms an out-of-plane spin system.
In contrast, other R-element materials display either an in-plane or canted spin direction behavior \cite{venturini_magnetic_1991, yin_quantum-limit_2020, el_idrissi_magnetic_1991, kimura_high-field_2006, ma_rare_2021, lee_interplay_2022}.
In particular, this case is interesting because the Chern magnetic phase predicted by Haldane's model requires an out-of-plane spin alignment \cite{haldane_model_1988}.
Initially, Haldane considered a honeycomb lattice \cite{haldane_model_1988}; however, the model was later extended to kagome systems \cite{ohgushi_spin_2000}.
Therefore, the kagome lattice with out-of-plane spin alignment requirement seems to be met by Mn atoms of TbMn$_6$Sn$_6$ and therefore opens the possibility of realizing a Chern magnet in this material.

Indeed, a few years ago, Yin et al.~\cite{yin_quantum-limit_2020} carried out experimental transport measurements using scanning tunneling microscopy (STM) in high-magnetic fields and observed a series of Landau quantization states in TbMn$_6$Sn$_6$.
They conducted AHE measurements and observed a large intrinsic contribution due to the large Berry curvature of Chern-gapped Dirac fermions.
Further edge-state conductivity measurements showed significant signals within the Chern gap energy scale.
Therefore, they concluded that TbMn$_6$Sn$_6$ is close to realizing a quantum-limit Chern magnet, as predicted by the Haldane model.
Indeed, kagome lattice geometry with an out-of-plane magnetization formed by the Mn atoms and the presence of strong spin-orbit coupling (SOC) stemming from the Tb and Sn atoms to open a Chern gap seems to suggest that TbMn$_6$Sn$_6$ could realize such an exotic phase.

However, the possibility of realizing the Chern phase hinges on the ability to align the Fermi level ($E_\mathrm{F}$) in the Dirac crossing (DC), or the Chern gap, seen in the kagome systems \cite{ohgushi_spin_2000, yin_quantum-limit_2020, lee_interplay_2022}.
Yin and coworkers' experiments show that the Chern gap is located at $130(4)$~meV above $E_\mathrm{F}$ \cite{yin_quantum-limit_2020}.
However, Lee and coworkers reported density functional theory with Hubbard $U$ (DFT$+U$) calculations~\cite{lee_interplay_2022} predicting the quasi-2D DC to be 700 meV above $E_\mathrm{F}$, or at least 300 meV above $E_\mathrm{F}$, suggesting that the Dirac crossing cannot be related to the observed AHE signal.
Therefore, there seems to be a disagreement in the literature about the exact position of DC with respect to $E_\mathrm{F}$ and whether TbMn$_6$Sn$_6$ could realize a Chern phase.
In this work, we focus on whether the DC is indeed close to the Fermi level using quantum Monte Carlo (QMC), DFT$+U$, and other correlated methods.

Clearly, accurately estimating the energetic level of DC with respect to $E_\mathrm{F}$, which is on the order of a few hundred meV, would require a very accurate electronic structure treatment.
Such a small energetic level could be affected by the accuracy of pseudopotentials \cite{hegde_quantifying_2022}, the level of electron correlation treatment, the accuracy of SOC effects, and 
the degree of multi-reference character intrinsic to the true electronic wave function of the system.
In the following sections, we thoroughly investigate these aspects and try to enumerate their contributions and biases.

In Section \ref{sec:pseudo}, we optimize the required Tb and Sn element pseudopotentials using correlated methods and evaluate their errors via transferability tests.
Then in Section \ref{sec:exact}, we carry out nearly-exact energy calculations for related molecular systems to estimate the biases present in single-reference QMC calculations.
We then apply the developed pseudopotentials to TbMn$_6$Sn$_6$ using DFT$+U$ in Section \ref{sec:dfa_sens} and QMC in Section \ref{sec:qmc} and demonstrate the issues with single-reference calculations. 
Section \ref{sec:dft_u} includes additional DFT$+U$ results, which elucidate the position of DC.
The findings are further analyzed and discussed in Section \ref{sec:discus}.


\section{Results}
\label{sec:results}

\subsection{Pseudopotentials}
\label{sec:pseudo}

One of the major approximations used in correlated methods is the pseudopotential or effective core potential (ECP) approximation \cite{dolg_relativistic_2012}.
High-accuracy correlated methods such as configuration interaction (CI) or coupled cluster (CC) are limited by the accuracy of the ECP as it modifies the Hamiltonian.
Therefore, it's crucial that ECPs be optimized in a correlation-consistent way, especially for correlated methods and systems.
Here, we develop correlation-consistent ECPs (ccECP) for Sn and Tb elements using CC and CI methods and use a ccECP for Mn from a previously published work \cite{kincaid_correlation_2022}.
In the next subsection, we show the results for the Tb element and compare its accuracy against fully correlated, relativistic, all-electron (AE) calculations.
The full results for Sn, optimization methods, and further details are provided in Supplementary Note 2.

\subsection{Tb ccECP}

The optimization of the Tb ECP is more involved than Sn (Supplementary Note 2) or Mn \cite{kincaid_correlation_2022}.
For instance, in Sn, the electrons can be easily partitioned by quantum principle number $n$ into core and valence spaces by treating $n \geq 5$ in the valence space ($5s$, $5p$, $5d$, ...).
However, in Tb, core$\leftrightarrow$valence partitioning is not straightforward since $4f$ orbital eigenvalues are on the same order as $n=5$ principal number orbitals ($5s$, $5p$, $5d$), suggesting it should be included in the valence space, see Supplementary Tables 7 and 8.
To overcome this, one could include the whole $n=4$ electrons in the valence and treat only $n \leq 3$ electrons as the core; however, this results in a prohibitively high computational cost in many-body QMC calculations.
Conversely, $4f$ electrons are sometimes included in the core, but this can result in poor results \cite{hegde_quantifying_2022}.
In this work, we decided to include the $4f$ electrons in the valence (with $N_\mathrm{core} = 46$ $e^-$) and treat them fully self-consistently along with $5s$, $5p$, $5d$, and $6s$ orbitals.
However, the many-body accuracy of such partitioning is not well-known and must therefore be carefully tested.
At least in $3d$ transition metals, similar large-core ECPs with valence spaces of $3d^k 4s^2$ (where the lowest $n$ orbital is not an $s$) are well-known to produce inaccurate results \cite{lee_quantum_2004, koseki_quantum_2008, burkatzki_energy-consistent_2008}.


\begin{figure*}[!htbp]
\centering
\includegraphics[width=\textwidth]{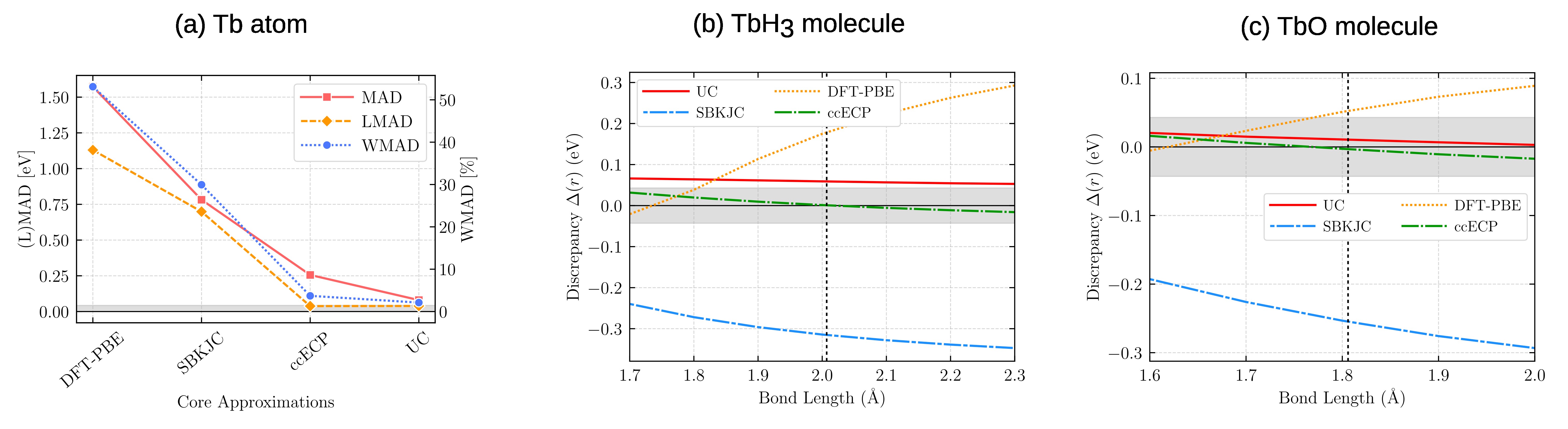}
\caption{
Scalar relativistic AE UCCSD(T) errors for various Tb core approximations.
(a) Tb atomic gap errors.
Various mean absolute deviation (MAD) metrics are used (see Supplementary Note 2 for details).
(b) TbH$_3$ and 
(c) TbO molecular binding energy discrepancies.
The shaded region indicates the band of chemical accuracy.
The dashed vertical line represents the equilibrium geometry.
In TbH$_3$, only the bond lengths are varied while the H$-$Tb$-$H angles of $120^{\circ}$ are kept constant.
The Tb atomic energy reference is [Xe] $4f^{8}6s^{2}5d^{1}$ state.
}
\label{fig:Tb_errors}
\end{figure*}



Our extensive many-body calculations show that the chosen core choice for Tb ccECP should result in small errors once properly optimized.
In fact, the accuracy of Tb ccECP is on par with the uncorrelated core all-electron (UC) where $N_\mathrm{core} = 46$ electrons are not active.
This is illustrated by the atomic gap errors in Figure \ref{fig:Tb_errors}(a) and TbH$_3$, TbO molecular binding energy errors in Figure \ref{fig:Tb_errors}(b-c).
We observed that all atomic gap errors are below $\sim$0.07 eV up to the third ionization potential level, see Supplementary Table 9.
Higher-order ionizations result in significantly larger errors on the order of an eV, showing the limits of such an ECP.
On the other hand, very high ionization potential errors are probably not relevant for condensed matter applications since rare-earth elements form either R$^{2+}$ or R$^{3+}$ compounds \cite{mugiraneza_tutorial_2022}.
Nevertheless, Tb ccECP is much more accurate than the existing ECPs with the same core choices, such as SBKJC \cite{cundari_effective_1993} and an \textsc{Opium}-generated DFT-PBE \cite{opium_code} ECP using the Troullier-Martins scheme \cite{troullier_efficient_1991}.
In fact, the ccECP errors are below chemical accuracy for the low-lying states (Figure \ref{fig:Tb_errors}(a)) and most of the bond lengths in hydride and oxide molecules (Figure \ref{fig:Tb_errors}(b-c)).
Tb and Sn ccECPs are available online in Ref \cite{pseudopotential_library} for broader use.

The presented low biases of Tb and Sn ccECPs here and of Mn ccECP elsewhere \cite{kincaid_correlation_2022} provide the opportunity to focus on the other systematic biases, such as the accuracy of the wavefunction and correlation treatment.
Namely, we can assume that any deviations from the experiments are not due to the technical limitations of the representative Hamiltonians; rather, it has to be due to more fundamental reasons, such as inaccurate trial wave function in QMC or improper density functional approximation (DFA) and Hubbard $U$ in DFT$+U$.
These are the subjects of Sections \ref{sec:exact} and \ref{sec:dfa_sens}, respectively.

\subsection{QMC Biases of Elemental Sub-Species}
\label{sec:exact}

It is beneficial to know the inherent systematic biases of single-reference QMC calculations for the system of interest by comparing the QMC energies to CI or CCSD(T) energies at the complete basis set limit (CBS).
Although CI and CCSD(T) calculations have been recently applied with great success to periodic systems \cite{gruber_applying_2018, benali_toward_2020, gallo_periodic_2021, wang_excitons_2020, mihm_shortcut_2021, neufeld_ground-state_2022, gao_electronic_2020}, they are still out of reach for large systems such as TbMn$_6$Sn$_6$ \cite{benali_toward_2020, gao_electronic_2020}.
Therefore, we focus on constituent atoms and relevant molecular systems to estimate the QMC biases.
Once the ECPs are made accurate, and other systematic errors such as timestep bias and walker population biases are under control, the main remaining errors are fixed-node and localization bias (if a nonlocal ECP is used as here).
Usually, the fixed-node bias is the dominant of these two \cite{mitas_pseudopotential_1993, krogel_magnitude_2017, dzubak_quantitative_2017}; however, both of them go to zero as the trial wave function approaches the exact form \cite{foulkes_quantum_2001, krogel_magnitude_2017, clay_deuterium_2019}.



\begin{figure*}[!htbp]
\centering
\includegraphics[width=0.75\textwidth]{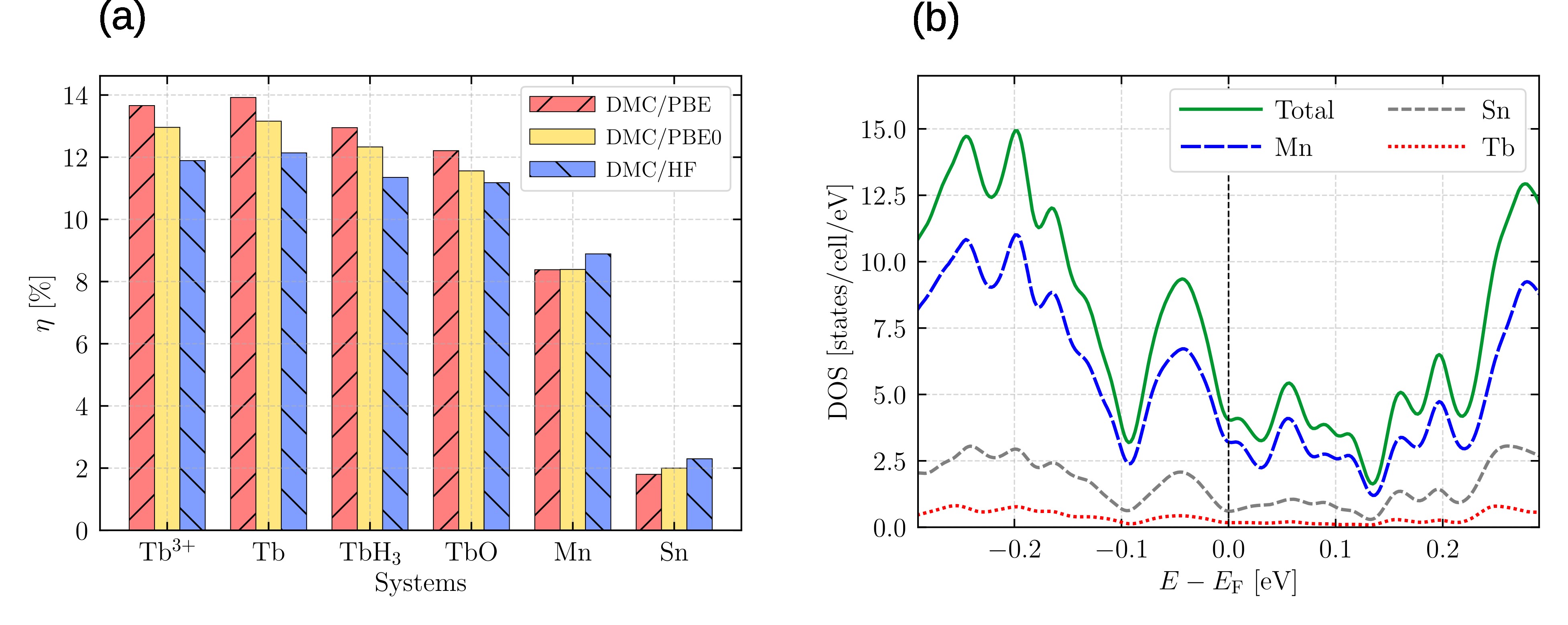}
\caption{
Element-resolved sources of errors. 
(a) Estimation of combined FN and localization biases in single-reference DMC calculations for TbMn$_6$Sn$_6$ relevant systems.
Here $\eta$ represents the missing percentage of the correlation energy in such calculations, $\eta = (E_\mathrm{exact} - E_\mathrm{DMC}) / (E_\mathrm{exact} - E_\mathrm{ROHF}) \cdot 100\%$.
(b) Density of states of bulk TbMn$_6$Sn$_6$.
Note that Mn atomic states dominate in this small energy window while Tb states are minimal.
LDA$+U^{3d}_\mathrm{Mn}(-0.5$ eV) DFA was used.
}
\label{fig:error_sources}
\end{figure*}

We enumerate the biases of fixed-node diffusion Monte Carlo (DMC) \cite{foulkes_quantum_2001} $\eta$ as a percentage of the correlation energies in Figure \ref{fig:error_sources}(a) using nearly-exact energies from CCSD(T) at CBS, and single-reference DMC calculations with PBE \cite{perdew_generalized_1996}, PBE0 \cite{perdew_rationale_1996}, and Hartree-Fock (HF) references. 
This plot reveals a few insights about TbMn$_6$Sn$_6$ solid.
First, the Sn atom has the smallest fixed-node bias, $\sim 2\%$ of the correlation energy.
This error is close to the $1 - 2 \%$ errors of the iso-valent Si atom, Si$_x$H$_y$ molecules, and diamond-structure bulk Si observed before \cite{wang_binding_2020, annaberdiyev_cohesion_2021}.
Second, Mn shows a sizable error, $\sim 9\%$ of the correlation energy missing in single-reference calculations, which also agrees with previous work \cite{annaberdiyev_accurate_2020}.
Finally, the largest errors occur for systems with Tb, where $\sim 12\%$ of the correlation energy is missing.
However, we found that these large biases tend to cancel out for Tb when considering energy differences, such as binding energies of TbH$_3$ and TbO molecules (see Supplementary Table 17).
%

In TbMn$_6$Sn$_6$, the leading Tb atomic density of states (DOS) is well below the Fermi level ($E_\mathrm{F}$), and Mn atomic DOS is dominant near $E_\mathrm{F}$ as shown in Figure \ref{fig:error_sources}(b).
Given that the Sn atom DMC errors are small and Tb states are well below $E_\mathrm{F}$, we argue that the largest errors in QMC will stem from an improper characterization of Mn atoms in our trial wave functions.
In other words, low-lying conduction states near $E_\mathrm{F}$ might give rise to wave functions of multi-reference character due to sizable single-reference DMC error in the Mn atom.
This is supported by calculated atomic moments from DFT$+U$, which show severe sensitivity when $U$ is applied on Mn-$3d$ orbitals (shown later) and insensitivity on Tb-$4f$, Tb-$5d$, and Sn-$5p$ orbitals, see Supplementary Table 15.
In addition, we observe no changes in the band structure when $+U$ is applied on Tb-$4f$ orbitals (Supplementary Figure 3).


Another insight from Figure \ref{fig:error_sources}(a) is that for a given magnetization, Tb systems with HF reference result in the lowest energies signifying the importance of localization in $4f$ orbitals.
On the other hand, Mn and Sn seem to favor more delocalized orbitals when compared to HF for constrained magnetization.
These aspects will be discussed later in detail by means of applying Hubbard $U$ on Mn-$3d$ orbitals.

\subsection{DFA Sensitivity Problem}
\label{sec:dfa_sens}

This work uses the DFT$+U$ approach with an effective Hubbard repulsion ($U_\mathrm{eff} = U - J$) \cite{cococcioni_linear_2005, dudarev_electron-energy-loss_1998} with the fully localized limit (FLL) double counting scheme \cite{anisimov_density-functional_1993, anisimov_first-principles_1997, liechtenstein_density-functional_1995, dudarev_electron-energy-loss_1998} as implemented in \textsc{quantum espresso} code \cite{giannozzi_quantum_2009, giannozzi_advanced_2017, giannozzi_quantum_2020}.
We explored two types of DFT$+U$ calculations.
In one case, the total magnetization of the cell was constrained to agree with the experimentally observed value of $\sim M^\mathrm{s} = 7~\mu_\mathrm{B}$ in the SCF procedure (indicated with C-DFT, such as C-LDA or C-PBE).
In the other case, the total magnetization was unconstrained and determined during the SCF cycles (simply indicated with the DFA name, such as LDA or PBE).
Extended details of the employed DFT$+U$ methodology are given in Supplementary Note 3.


\begin{figure*}[!htbp]
\centering
\includegraphics[width=\textwidth]{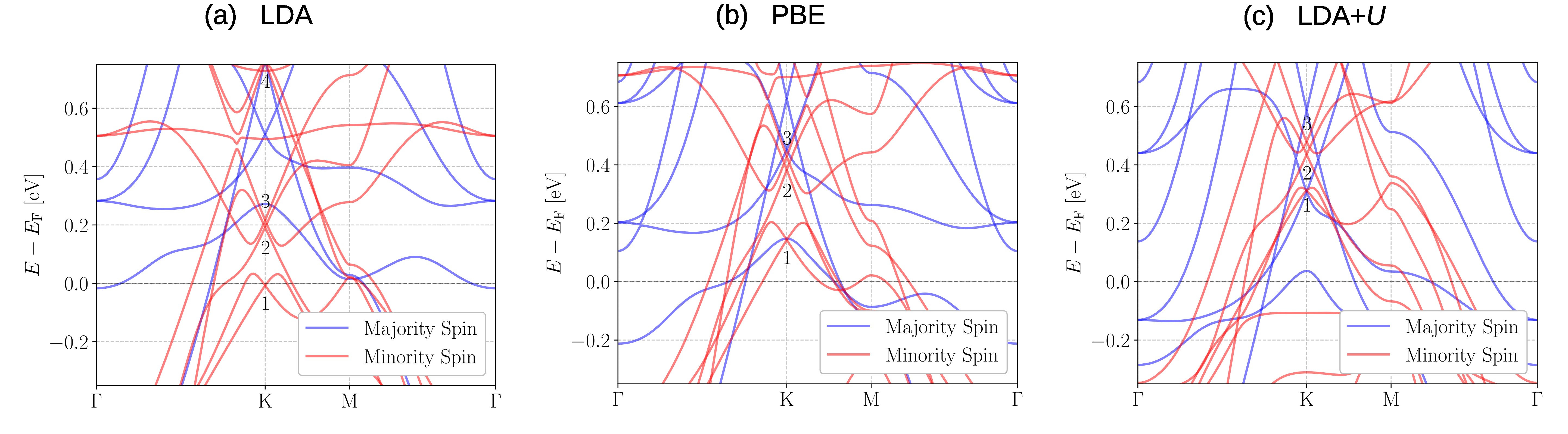}
\caption{
Band structures of TbMn$_6$Sn$_6$ using scalar relativistic (averaged spin-orbit) ECPs.
DFAs employed here are
(a) LDA,
(b) PBE,
(c) LDA with $+U^{3d}_\mathrm{Mn}(+0.5$ eV).
Note how severely DC positions (labeled by numbers at $k=$ K) shift as the DFA is changed.
}
\label{fig:bulk_bands}
\end{figure*}

TbMn$_6$Sn$_6$ ground state is a ferrimagnetic phase (FiM), where the Mn atoms align ferromagnetically, and the Tb spins point in the opposite direction \cite{el_idrissi_magnetic_1991}.
Figure \ref{fig:bulk_bands} shows the bulk band structures using various DFAs for the FiM phase.
For a pure 2D kagome lattice, tight binding models show that the DC occurs at $E_\mathrm{F}$, $k =$~K point while the flat bands are above the $E_\mathrm{F}$ \cite{ohgushi_spin_2000, yin_quantum-limit_2020}.
Considering the LDA \cite{perdew_self-interaction_1981} band structure (Fig. \ref{fig:bulk_bands}(a)), we see the expected 'flat' band in the minority spin channel at $\sim$0.5~eV.
In addition, a few DCs occur at $k =$ K in the minority spin channel.
These DCs were labeled for clarity in Figure \ref{fig:bulk_bands} and will be referred to as DC$n$ where $n$ is the label.
When LDA is switched to PBE DFA (Fig. \ref{fig:bulk_bands}(b)), the flat band and DC2, DC3 shift up by around 200~meV.
Introducing a small effective Hubbard $U^{3d}_\mathrm{Mn} = 0.5$~eV to LDA (Fig. \ref{fig:bulk_bands}(c)) shifts the flat bands and DCs up even further.
The upward shifts in the minority spin channel are accompanied by downward shifts in the majority spin channel.
Obviously, this energetic sensitivity is a problem where one is interested in resolving an energy scale of $\sim 150$~meV.


\begin{figure*}[!htbp]
\centering
\includegraphics[width=0.75\textwidth]{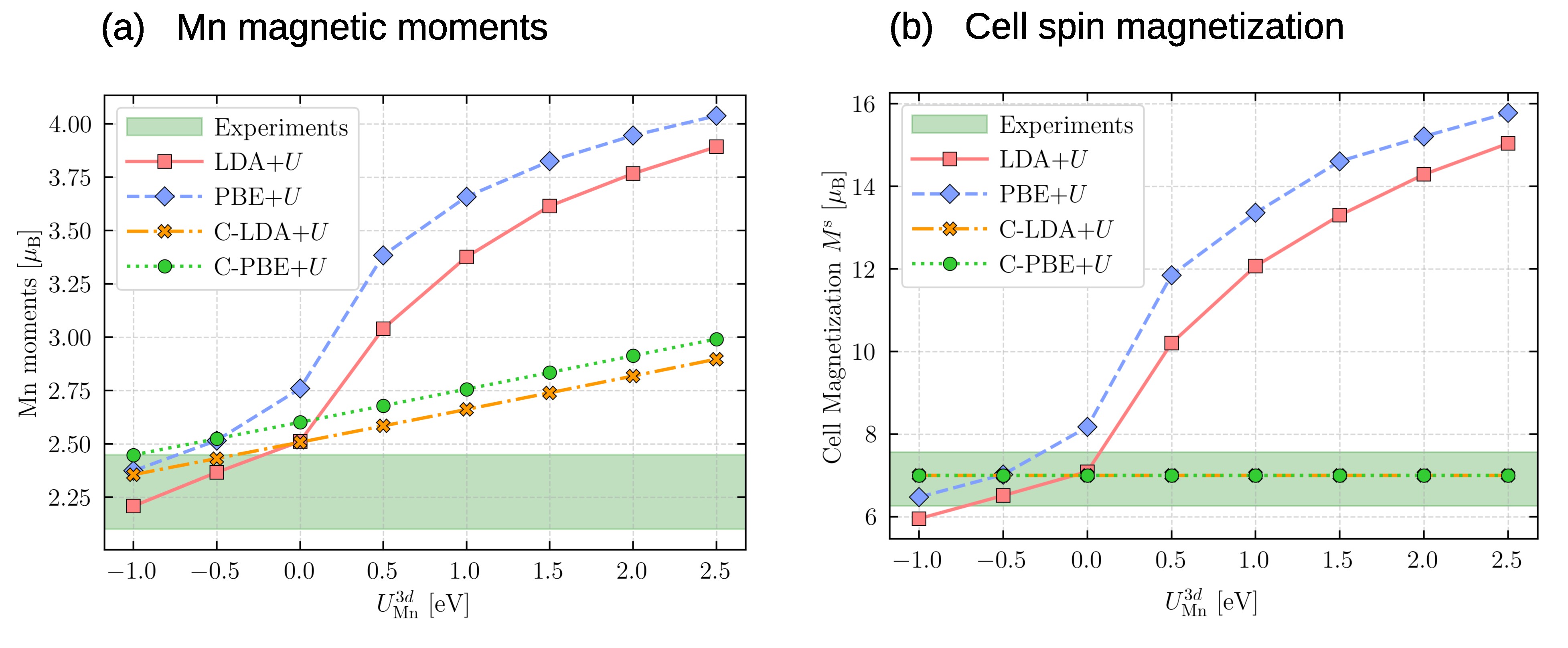}
\caption{
Comparison of DFT$+U^{3d}_\mathrm{Mn}$ and experimental magnetizations.
(a) Mn atomic magnetic moments using L\"owdin populations,
(b) cell spin magnetization per formula unit.
The green-shaded regions represent the envelope of various experimental measurements \cite{el_idrissi_magnetic_1991, venturini_magnetic_1991, clatterbuck_magnetic_1999, mielke_iii_low-temperature_2022, kimura_high-field_2006, yin_quantum-limit_2020, jones_origin_2022}.
In (b), the experimental cell magnetization was added with the Tb orbital magnetization of $M^\mathrm{orb}_\mathrm{Tb} = 2.96~\mu_\mathrm{B}$ \cite{lee_interplay_2022} for a proper comparison, $M^\mathrm{s} = M^\mathrm{total}_\mathrm{exp} + M^\mathrm{orb}_\mathrm{Tb}$.
}
\label{fig:mag_exp_vs_theory}
\end{figure*}

Another related issue is demonstrated in Figure \ref{fig:mag_exp_vs_theory} by considering the Mn magnetic moment and cell magnetization as $U^{3d}_\mathrm{Mn}$ is varied.
Strikingly, the magnetizations are severely overestimated compared to the experimental green-shaded region once a small positive value of $U^{3d}_\mathrm{Mn}$ is introduced.
In fact, the overestimation of magnetic moments with advanced DFAs or DFT$+U$ is a well-known issue for metals displaying itinerant magnetism \cite{fu_applicability_2018, fu_density_2019, lee_interplay_2022, ghosh_unraveling_2022}.
Previous studies \cite{fu_applicability_2018, fu_density_2019} showed that increasingly advanced DFAs in Jacob's ladder of DFA ranks \cite{perdew_jacobs_2001} such as GGA PBE \cite{perdew_generalized_1996}, meta-GGA SCAN \cite{sun_strongly_2015}, and hybrid PBE0 \cite{perdew_rationale_1996} tend to increasingly overestimate the magnetic moments in metallic ferromagnetic systems such as bcc Fe, hcp Co, and fcc Ni.
Interestingly, even LDA could slightly overestimate the magnetic moments, such as in fcc Ni \cite{fu_density_2019}.

In Figure \ref{fig:mag_exp_vs_theory}, it is evident that LDA$+U$ requires a small negative $U^{3d}_\mathrm{Mn}$ value to agree with both experimental Mn atomic moment and overall cell magnetization.
We note that utilization of a negative $U$ was suggested \cite{kulik_perspective_2015, cococcioni_lda_2012} and applied \cite{persson_improved_2006, nakamura_lda_2008, nakamura_first-principle_2009, micnas_superconductivity_1990, hase_madelung_2007} before, especially on superconducting systems.
Physically, this means that Hund's exchange $J$ dominates the on-site Coulomb repulsion $U$, resulting in a minor negative $U_\mathrm{eff} = U - J$ value.
The result is more delocalization of orbitals and a tendency to pair electrons, which produce lower magnetic moments.
The opposite extreme can be seen when using HF and hybrid DFA for metals, which over-localize the electrons and tend to unpair the electrons, resulting in overestimated magnetic moments.
Even though DFT$+(U > 0)$ physics is different than hybrid DFAs, the effect is a similar localization of orbitals \cite{kulik_perspective_2015}.
In this case, the proper delocalization could be achieved by a minor negative $U_\mathrm{eff}$ or a Hund's $J$ with a magnitude larger than $U$.
In addition, previous studies showed that the employed double-counting scheme in DFT$+U$ could significantly change the predicted magnetic moments \cite{lee_interplay_2022, ryee_effect_2018}.
Ultimately, the key goal is to achieve a proper (de)localization of electrons and hence the correct magnetization driven by an appropriate tendency of electrons to pair.
The difficulties of predicting magnetic moments using advanced DFAs raise the issue of whether single-reference QMC calculations can predict the correct magnetization in a complex correlated metal such as TbMn$_6$Sn$_6$.

\subsection{Single-Reference QMC}
\label{sec:qmc}

\begin{figure}[!htbp]
\centering
\includegraphics[width=0.5\textwidth]{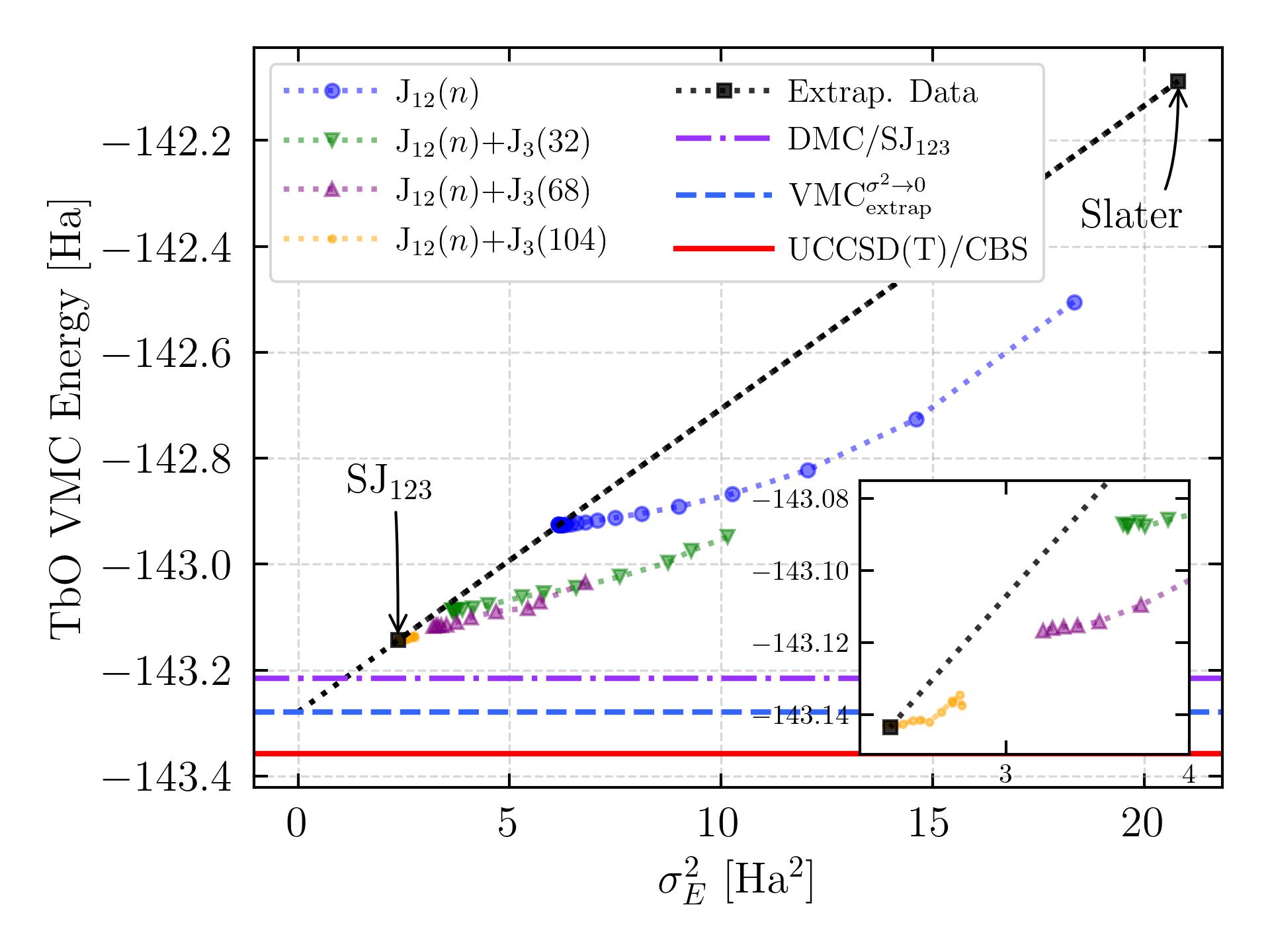}
\caption{
VMC energy vs. variance extrapolation for TbO molecule.
DMC/SJ$_{123}$ is a DMC calculation using HF Slater-Jastrow form with one-, two-, and three-body terms in the Jastrow.
UCCSD(T)/CBS represents the nearly-exact energy from basis set extrapolations.
In J$_{12}(n)$, $n = (16,...,100)$.
All data points were optimized for energy.
See the main text for details.
}
\label{fig:vmc_TbO}
\end{figure}

As kinetic energy sampling via $k$-mesh integration is crucial in metals \cite{azadi_systematic_2015, azadi_efficient_2019, dagrada_exact_2016, lin_twist-averaged_2001}, we use canonical twist-averaging (CTA) and a supercell with two formula units (Figure \ref{fig:qmc_cell}) in all TbMn$_6$Sn$_6$ QMC calculations.
To assess the validity of single-reference QMC, we consider results from two different QMC methods.
These methods are fixed-node/fixed-phase DMC (captures most correlations), and \mbox{$\sigma^2 \rightarrow 0$} extrapolated VMC ({VMC}$^{\sigma^2 \rightarrow 0}_{\rm extrap}$, estimate of converged dynamic correlations) using single-reference Slater-Jastrow type trial wave functions.
The {VMC}$^{\sigma^2 \rightarrow 0}_{\rm extrap}$ method uses two data points with the same determinantal form to extrapolate to the 
\mbox{$\sigma^2=0$} limit as shown in Figure \ref{fig:vmc_TbO} for TbO molecule where nearly-exact energies can be obtained via UCCSD(T)/CBS.
Specifically, energies from the Slater trial wave function and Slater-Jastrow trial wave function with one-, two-, and three-body terms (SJ$_{123}$) were used in the extrapolation (black squares).
In principle, as the variance goes to zero, the VMC energy should recover the exact correlation energy due to zero-variance principal \cite{foulkes_quantum_2001}.
However, since the Jastrow factor only captures the dynamic correlations \cite{huang_accuracy_1997}, we do not expect this estimator to recover the exact correlation, as the trial wave function is improved only in the symmetric term (due to Jastrow), while the antisymmetric term (due to Slater) remains as an approximation.
Although one can not strictly distinguish between static and dynamic correlations, the above separation is clear -- the Jastrow term is the source of 'dynamic correlations', while the fermionic term is the source of 'static correlations'.
In this sense, the {VMC}$^{\sigma^2 \rightarrow 0}_{\rm extrap}$ method is expected to give the energy for the 'perfect Jastrow'.
We would like to note that this is only an approximation, as \mbox{$\sigma^2=0$} should only occur for the exact ground state.

Provided that the trial wave function $\Psi_T$ has a significant overlap with the exact ground state only, one can show that the VMC energy must linearly depend on the variance for $\sigma^2 \rightarrow 0$ \cite{taddei_iterative_2015}:
\begin{equation}
    E_{\mathrm{VMC}} = E_{\mathrm{exact}} + \alpha \sigma^2.
\end{equation}
In practice, it can be difficult to achieve near-ideal linear extrapolations, for instance, as seen in multi-determinant extrapolations of high-pressure hydrogen \cite{clay_deuterium_2019}.
We investigated the convergence behavior of one- and two-body Jastrow (J$_{12}(n)$) and also three-body Jastrow (J$_3(n)$) as the number of optimizable parameters $n$ is increased.
A nonlinear convergence was observed for J$_{12}(n)$ in all cases of Figure \ref{fig:vmc_TbO}.
For large $n$, J$_{12}(n)$ reaches a plateau where the energy is converged with respect to $n$.
A linear behavior was observed when using these converged J$_{12}$ functions and including J$_{3}(n)$ terms with various $n$ values.
The simple two-point extrapolations using the Slater trial wave function (free of Jastrow optimization imperfections) and the SJ$_{123}$ with the highest feasible optimizable parameters (the best Jastrow we could optimize) proved to be the most robust and consistent way to compare energies. 
The obtained {VMC}$^{\sigma^2 \rightarrow 0}_{\rm extrap}$ value would therefore represent the VMC energy of a SJ$_{(123...\infty)}$, i.e, Jastrow with up to $M$-body interactions where $M\rightarrow\infty$.

In Figure \ref{fig:vmc_TbO}, even though {VMC}$^{\sigma^2 \rightarrow 0}_{\rm extrap}$ estimator energy is below the single-reference DMC, it is still significantly above the estimated exact energy since the estimator is mainly probing for the effect of dynamic correlations due to single-reference form.
We observed similar plots with the same energy ordering for the other molecules considered in this work; see Supplementary Figure 14.
The only exception to this was the Sn atom, where the {VMC}$^{\sigma^2 \rightarrow 0}_{\rm extrap}$ estimated energy goes slightly below the UCCSD(T)/CBS value (by about $10^{-4}$~Ha).
This is not surprising since the fixed-node bias in the Sn atom is very small ($\sim 2\%$ of the correlation energy), and thus we might expect that the static correlation is captured well by single-reference and {VMC}$^{\sigma^2 \rightarrow 0}_{\rm extrap}$ should give close to exact energies.
More information about VMC energy extrapolations and extended details of QMC methods are given in Supplementary Note 4.

In Figure \ref{fig:QMC_scan}(a-b), we plot the bulk energies with the above-described methods where the orbitals are generated from DFA = \{LDA, PBE\}.
Specifically, LDA DFT calculations obtain magnetizations close to the experiments, so it is interesting to see whether QMC predicts the correct magnetization when the LDA orbitals are reused for various QMC magnetizations.
Namely, in Figure \ref{fig:QMC_scan}(a-b), the same orbitals are used, but a different overall magnetization $M^\mathrm{s}$ is constructed in QMC calculations. 
Since QMC methods are variational, the lowest energy states correspond to the physical predictions from single reference QMC \cite{kylanpaa_accuracy_2017, wines_systematic_2023}.
As shown in these plots, the lowest energy for both estimators corresponds to $\sim 15~\mu_\mathrm{B}$, approximately twice the correct value and well outside of the experimental region.
This overestimation seems independent of the employed orbitals, as LDA and PBE results follow similar trends.
In addition, extrapolating the dynamic correlations to the zero-variance limit does not change the energy curve, signifying a necessity for either significantly improved orbitals or including static correlations via multi-reference wavefunctions.


\begin{figure*}[!htbp]
\centering
\includegraphics[width=0.75\textwidth]{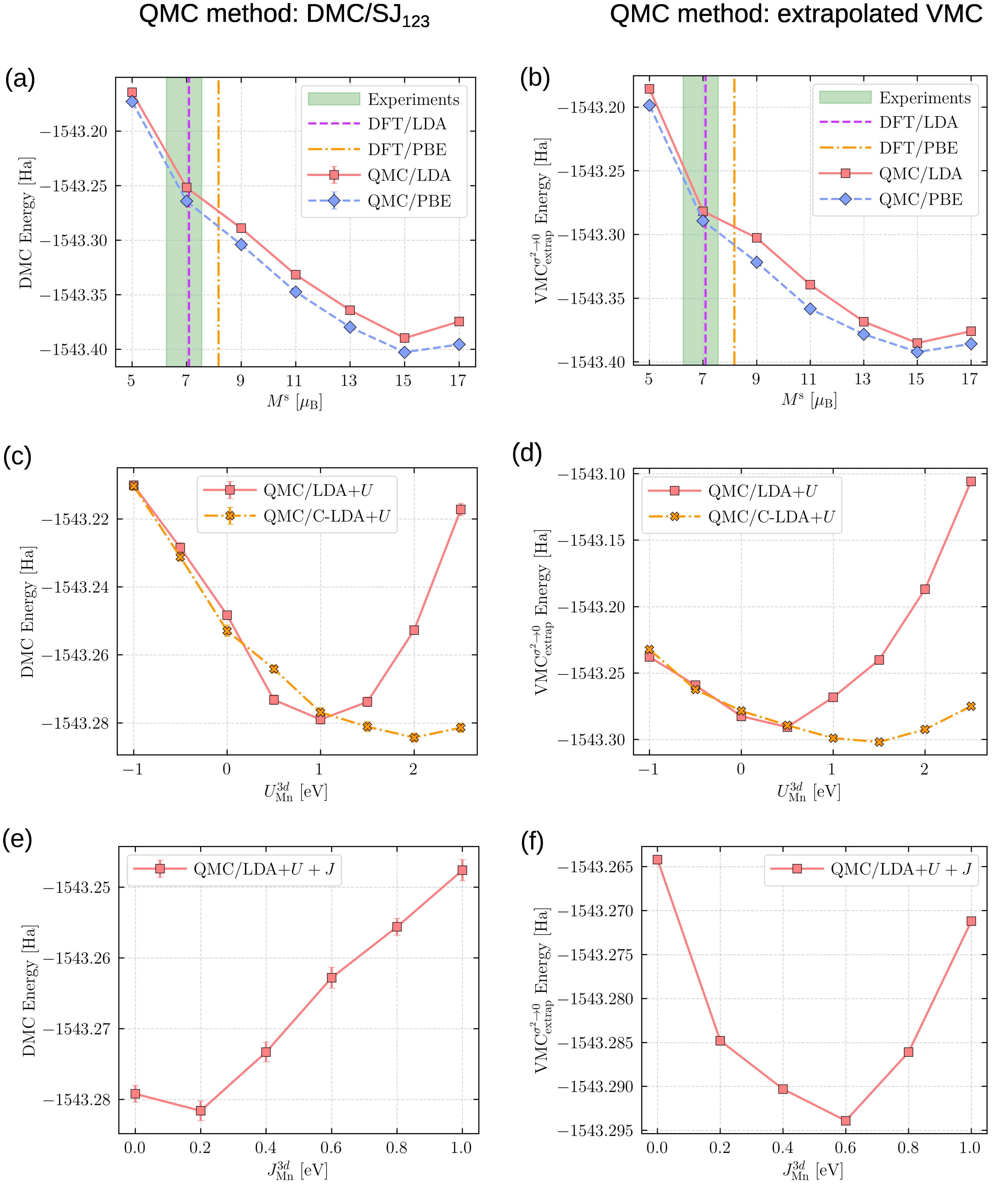}
\caption{
QMC energies of various trial wave functions.
(a-b) QMC energies [Ha] using LDA and PBE trial wave functions for various QMC magnetizations $M^\mathrm{s}$.
The green-shaded regions represent the envelope of various experimental measurements \cite{el_idrissi_magnetic_1991, venturini_magnetic_1991, clatterbuck_magnetic_1999, mielke_iii_low-temperature_2022, kimura_high-field_2006, yin_quantum-limit_2020, jones_origin_2022}.
The experimental cell magnetization was added with the Tb orbital magnetization of $M^\mathrm{orb}_\mathrm{Tb} = 2.96~\mu_\mathrm{B}$ \cite{lee_interplay_2022} for a proper comparison, $M^\mathrm{s} = M^\mathrm{total}_\mathrm{exp} + M^\mathrm{orb}_\mathrm{Tb}$.
DFT-predicted magnetizations are also shown as vertical lines.
(c-d) QMC energies [Ha] using DFT$+U_\mathrm{eff}$~\cite{cococcioni_linear_2005, dudarev_electron-energy-loss_1998} and (e-f) DFT$+U+J$ \cite{liechtenstein_density-functional_1995} trial wave functions.
The QMC spins in (c-f) are constrained at $M^\mathrm{s} = N_{e}^\mathrm{up} - N_{e}^\mathrm{down} = 7~\mu_\mathrm{B}$ for each twist.
QMC error bars correspond to one standard deviation and are either similar to or smaller than data symbols.
}
\label{fig:QMC_scan}
\end{figure*}

Next, we explore fixing the QMC cell magnetization to agree with experiments (at $M^\mathrm{s} = 7~\mu_\mathrm{B}$) and varying the orbitals by the introduction of Hubbard $+U$ (Figure \ref{fig:QMC_scan}(c-f)).
In Figures \ref{fig:QMC_scan}(c-d), this is done for unconstrained DFT ($M^\mathrm{s}$ from SCF) and magnetization-constrained DFT ($M^\mathrm{s} = 7~\mu_\mathrm{B}$) using an effective Hubbard $U$ as described in Section \ref{sec:dfa_sens}.
In Figure \ref{fig:QMC_scan}(c), the DMC method obtains the lowest energy for unconstrained LDA at $U^{3d}_\mathrm{Mn} = 1$~eV.
The magnetization in this reference is severely overestimated in DFT with cell magnetization of $\approx 12~\mu_\mathrm{B}$ and Mn moment of $3.38~\mu_\mathrm{B}$, see Figure \ref{fig:mag_exp_vs_theory}.
On the other hand, the lowest DMC energy for the constrained reference occurs at a much larger value of $U^{3d}_\mathrm{Mn} = 2$~eV.
This is because in the constrained case, the Mn moments increase much more slowly as $U^{3d}_\mathrm{Mn}$ is increased (Figure \ref{fig:mag_exp_vs_theory}(a)) requiring a larger value of $U^{3d}_\mathrm{Mn}$.
This suggests that the magnetization of Mn atoms dictates the QMC energy.

In Figure \ref{fig:QMC_scan}(d), similar plots are shown using {VMC}$^{\sigma^2 \rightarrow 0}_{\rm extrap}$ method.
The results are qualitatively similar to DMC in Figure \ref{fig:QMC_scan}(c), but the minima are shifted by about $\Delta U^{3d}_\mathrm{Mn} = 0.5$~eV closer to zero.
However, the corresponding DFT magnetizations are still considerably overestimated and outside the experimental measurements (see Figure \ref{fig:mag_exp_vs_theory} for corresponding $U^{3d}_\mathrm{Mn}$ values).
This shows that including the full dynamic correlations can improve the prediction of magnetic properties, but it is not enough for an accurate estimation in this case.

Finally, we explored the effect of $J$ in the DFT$+U+J$ formalism \cite{liechtenstein_density-functional_1995} in Figures \ref{fig:QMC_scan}(e-f).
We consider LDA with $U^{3d}_\mathrm{Mn} = 1.0$~eV, which results in the lowest energy in DMC for this DFA (Figure \ref{fig:QMC_scan}(c)).
In DMC/SJ$_{123}$ method, Figure \ref{fig:QMC_scan}(e), the energies increase as $J$ is increased.
This is because the effect of $J$ is to discourage high-spin states, and we see a similar effect as in Figure \ref{fig:QMC_scan}(c).
The VMC$^{\sigma\rightarrow 0}_{\rm extrap}$ in Figure \ref{fig:QMC_scan}(e) method predicts a value of $J^\mathrm{3d}_\mathrm{Mn} = 0.6$~eV; however, the corresponding Mn moment of $2.61~\mu_\mathrm{B}$ is still slightly overestimated relative to the experimental value of 2.39(8)~$\mu_\mathrm{B}$ \cite{el_idrissi_magnetic_1991} (see Supplementary Figure 8).


It is evident from Figure \ref{fig:QMC_scan}(c-f) that constraining the QMC cell magnetization seems to alleviate the overestimation problem.
This is especially pronounced in the VMC$^{\sigma\rightarrow 0}_{\rm extrap}$ case where full dynamic correlations are expected to be captured.
However, the lowest energies using constrained QMC cell magnetizations in Figure \ref{fig:QMC_scan}(c-f) are still much higher than the unrestricted, $M^\mathrm{s} = 15~\mu_\mathrm{B}$ case in Figure \ref{fig:QMC_scan}(a).
This is true within each method, DMC/SJ$_{123}$ and VMC$^{\sigma^2\rightarrow 0}_{\rm extrap}$.
These calculations show that single-reference QMC overestimates the magnetic moments similar to meta-GGA SCAN and hybrid PBE0 DFAs.
This overestimation trend persists even when the dynamic correlations are extrapolated to the zero-variance values, Figure \ref{fig:QMC_scan}(d).
This points to a significant deficiency in static correlations, which possibly stem from a few sources:

\begin{enumerate}
    \item Imperfect one-particle orbitals.  Within the single-reference framework, the nodal surface is fully determined by the single-particle inputs. In case of the presence of static/multi-reference correlations, this can bias the expectations. 
    Natural orbitals (NOs) from CI calculations can provide a better orbital set for improving the description within the single-reference model \cite{wang_performance_2019}.
    \item Lack of multi-reference trial wave function. Even with high-quality orbitals such as NOs, it is possible that multi-reference wave functions are needed to predict the magnetic moments correctly.
    This would not be surprising because systems with Mn elements are known to display multi-reference characters \cite{drosou_spin-state_2021}.
    For instance, a previous study of bulk MnO using CCSD found improvement in band gaps and magnetic moments on the overestimated values of UHF \cite{gao_electronic_2020}.

    An additional example where a similar problem occurs is the W atom.
    Its ground state is [Xe]$4f^{14} 5d^4 6s^2$ ($^5D$) configuration, while the first excited state configuration is a higher spin [Xe]$4f^{14} 5d^5 6s^1$ ($^7S$) \cite{NIST_ASD}.
    Fixed-phase spin-orbit DMC, with minimal expansion of determinants, incorrectly predicts the high-spin state $^7S$ as the ground state \cite{wang_new_2022}.
    However, as the trial wave function is expanded with more determinants, the correct $^5D$ ground state is recovered in agreement with the experiments \cite{wang_new_2022}.

    \item Finite-size effects.
    In this work, we used a supercell with two formula units and canonical twist-averaging where each twist is occupied with the same $M^\mathrm{s}$.
    Although CTA should reach the same TDL energy as grand-canonical twist averaging (GCTA) for the same magnetization, it could have a considerable impact on finite supercell sizes.
    This overlaps with the first point above as it is a one-body effect.
    However, it has a different origin since the change is across-twist rather than within-twist.
    
\end{enumerate}
Although the above effects can be present, it is out of the scope of this work to study TDL convergence and to apply multi-reference approaches here since we are interested in the single-reference model and its accuracy limit for these types of materials.
Therefore, we leave these aspects for future study.

Intuitively, the QMC overestimation issue could be explained as follows. 
Mn element in the atomic limit favors the $^6S_{5/2}$ high-spin state \cite{NIST_ASD}, and this localized atomic energy scale could dominate the QMC energy in the metallic solid.
The energy due to bonding is partially captured in the single-reference case; however, it is not properly balanced with the localized energy scale, which skews the optimal energy to the Mn atomic high-spin state.
Therefore, a balanced trial wave function would also need to capture the multi-reference effects stemming from crystal field and orbital hybridization so as to counterweight the pronounced atomic high-spin limit.
Indeed, in the HF method, where there are no correlations by definition, the Mn moment is close to the fully unpaired limit of $5~\mu_\mathrm{B}$, see Table \ref{tab:hybrids_mag}.
Compared to insulators, the interactions must be significantly screened in metals \cite{grosso_chapter_2014}, and HF or hybrid DFAs are well-known to result in poor screening \cite{kaxiras_2003, janeskoScreenedHybridDensity2009}, and other unphysical features in partially-filled band metals \cite{monkhorst_hartree-fock_1979, delhalle_direct-space_1987}.
We note that in the thermodynamic limit (TDL) of full-CI (FCI), the proper screening must be achieved in metallic systems due to the cancelation of static correlation effects and long-range interactions \cite{monkhorst_hartree-fock_1979, delhalle_direct-space_1987, janeskoScreenedHybridDensity2009}.
Remarkably, the approximate correlations in LDA seem to effectively capture the screening and multi-reference effects quite well, judging by the obtained magnetizations in Table \ref{tab:hybrids_mag}.
Indeed, a few previous studies found that LDA could mimic the long-range correlations due to multireference effects \cite{polo_long-range_2003, cremer_density_2001, kraka_dieter_2019}, which is facilitated by the larger self-interaction errors in LDA compared to hybrids.

\input{Table_1}


\subsection{DFT$+U$ Results}
\label{sec:dft_u}

In order to shed additional light on the intricate relationship of single-reference and magnetic order, we empirically adjust the magnetic moment of the Mn atom to the neutron diffraction experimental result of 2.39(8)~$\mu_\mathrm{B}$ \cite{el_idrissi_magnetic_1991}.
A recent muon spin rotation measurement, which is a highly powerful probe of local magnetism, also obtained an Mn moment of $\sim 2.4~\mu_\mathrm{B}$ at low temperatures \cite{mielke_iii_low-temperature_2022}.
We note that using the Mn moments as a reference is more robust than comparing against the overall cell magnetization or Tb atomic moments due to the orbital magnetic moment contribution of the localized and well-screened $4f$ orbitals \cite{mugiraneza_tutorial_2022}.
On the other hand, the $3d$ orbital magnetic moments are negligible due to orbital quenching \cite{mugiraneza_tutorial_2022, lee_interplay_2022}, and the spin moments can be directly compared with the experiments.
In addition, the Mn moments were shown to display much smaller spin fluctuations \cite{jones_origin_2022}, making Mn more suitable to compare against experiments.
In fact, using muon spin rotation experiments, Ref.~\cite{mielke_iii_low-temperature_2022} showed a critical slowing down of spin fluctuations below $T^*_\mathrm{C1} \simeq 120$~K.
Further colling the systems below $T_\mathrm{C1} \simeq 20$~K resulted in freezing the spin fluctuations into static patches of ideal out-of-plane FiM ordering, which persisted down to the lowest measured 1.7~K temperature.
Therefore, although our calculations are $T=0$~K and collinear FiM, they can be directly compared with the Mn moments obtained from $T < T_\mathrm{C1} \simeq 20$~K without considering the effects of spin fluctuations which are seen in some other frustrated kagome materials \cite{ghimire_competing_2020, lu_observation_2022, kolincio_kagome_2023}.
We find that employing LDA DFA and L\"owdin population analysis, a small effective value $U^{3d}_\mathrm{Mn} \approx -0.5$~eV is required to approximately match with the neutron diffraction experimental value of 2.39(8)~$\mu_\mathrm{B}$ \cite{el_idrissi_magnetic_1991} (LDA$+U^{3d}_\mathrm{Mn}(-0.5$~eV) results in Mn moment of $2.366~\mu_\mathrm{B}$).
This value of $U^{3d}_\mathrm{Mn}$ is qualitatively corroborated by Figure \ref{fig:error_sources}(a), where Mn seems to favor more delocalized orbitals for the given magnetization, and previous studies showing that even LDA, which presumably produces the smallest orbital localization, shows a tendency to overestimate the magnetic moments \cite{fu_applicability_2018, fu_density_2019}.

\begin{figure*}[!htbp]
\centering
\includegraphics[width=0.75\textwidth]{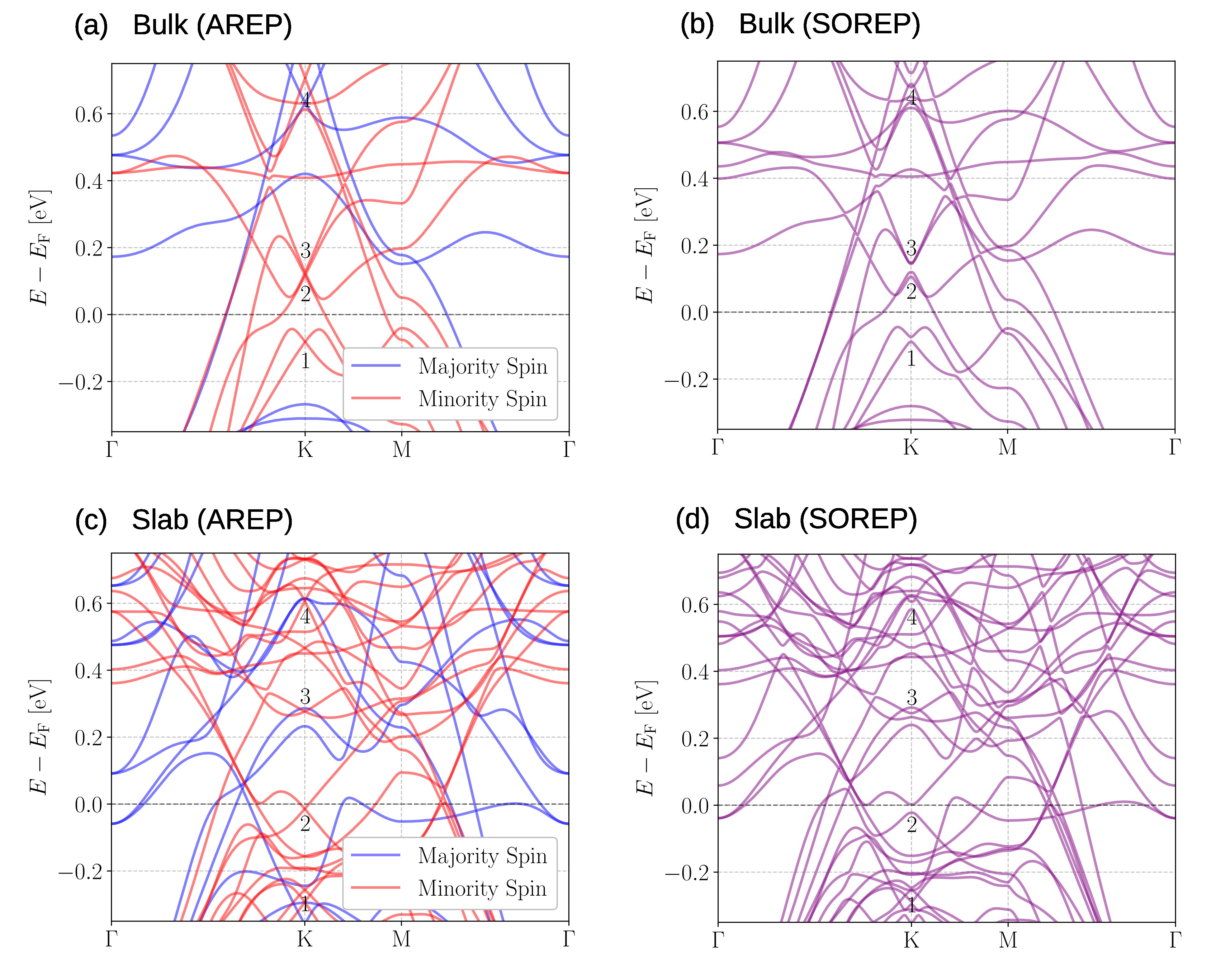}
\caption{
Band structures using LDA$+U^{3d}_\mathrm{Mn}$($-0.5$ eV) with scalar relativity (averaged spin-orbit coupling, AREP) as well as full relativity (explicit spin-orbit coupling, SOREP).
(a) Bulk with AREP,
(b) slab with AREP,
(c) bulk with SOREP,
(d) slab with SOREP.
Labels '1', '2', and '3' at $k=$ K indicate the DC positions in the minority spin channel.
}
\label{fig:dft_opt_bands}
\end{figure*}

Having established an appropriate DFA for the study of TbMn$_6$Sn$_6$, we now plot the bulk band structure for the FiM phase using LDA$+U^{3d}_\mathrm{Mn}$ with $U^{3d}_\mathrm{Mn} =-0.5$~eV.
In Figure \ref{fig:dft_opt_bands}(a), the bands are plotted with scalar relativistic pseudopotentials, namely, with averaged spin-orbit interactions.
We find that the DC2 energetic level ($E_\mathrm{DC2}$) with respect to $E_\mathrm{F}$ shifts down to about $E_\mathrm{DC2} = 120$~meV when compared with LDA (Figure \ref{fig:bulk_bands}(a)).
This is in excellent agreement with the experimental result of $E^\mathrm{exp}_\mathrm{DC} = 130(4)$~meV from tunneling and quasiparticle scattering along the bulk crystal edge direction \cite{yin_quantum-limit_2020}.
In addition, the inclusion of explicit SOC results in the opening of the Chern gap in DC2 and DC3 seen at 120 meV above $E_\mathrm{F}$, Figure \ref{fig:dft_opt_bands}(b).
The Chern gap for DC2 is $\Delta = 25$~meV, which is also in reasonable agreement with the experimental value of $\Delta^\mathrm{exp} = 34(2)$~meV.

\begin{figure}[!htbp]
\centering
\includegraphics[width=0.5\textwidth]{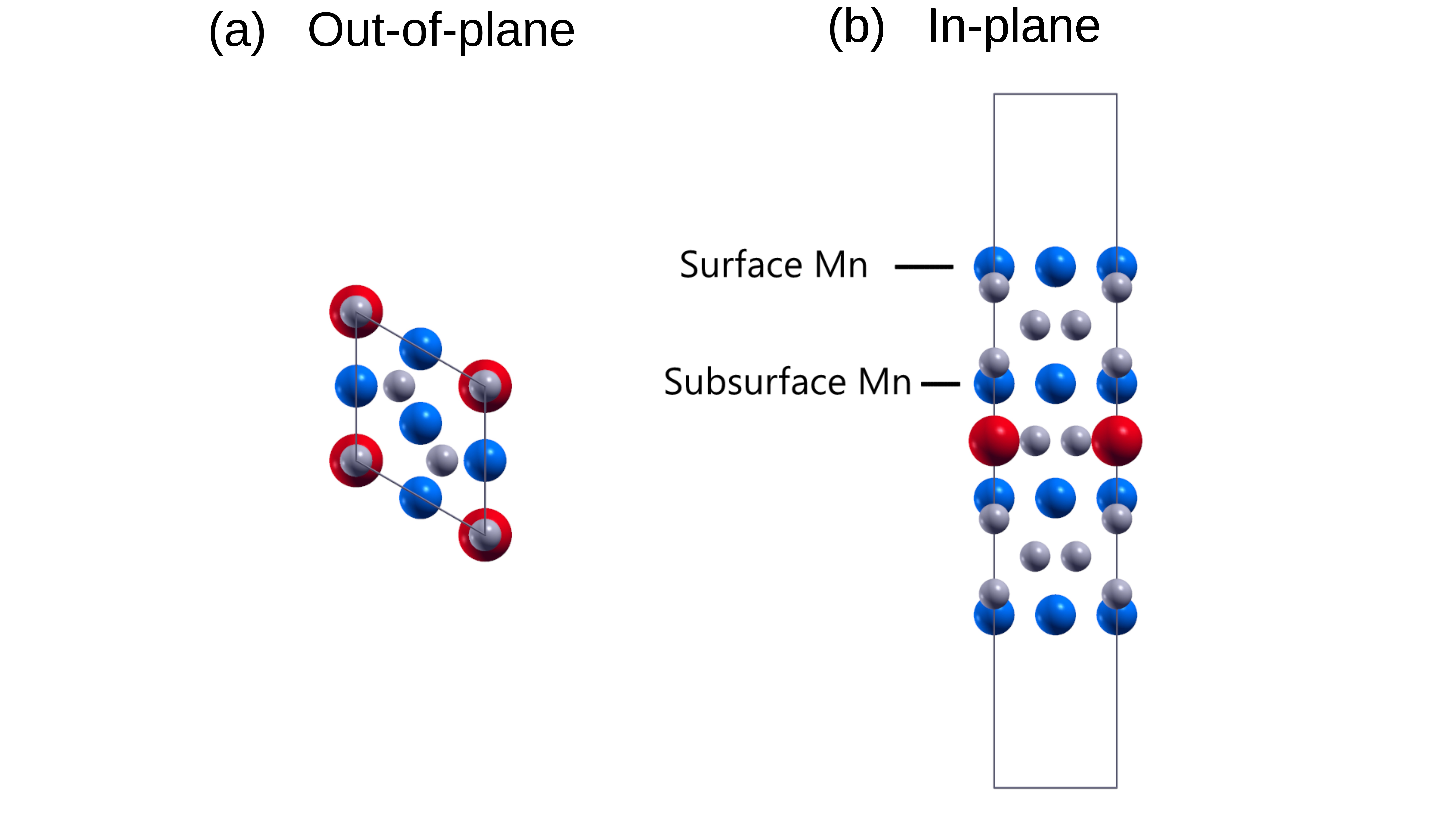}
\caption{
A slab of TbMn$_6$Sn$_6$.
(a) Out-of-plane and (b) in-plane views are shown.
The large red spheres are Tb atoms, medium blue spheres are Mn atoms, and small gray spheres are Sn atoms.
The slab is Mn-terminated due to the observed intense modulation in the experimental $\mathrm{d}I/\mathrm{d}V$ line map \cite{yin_quantum-limit_2020}.
The vacuum length between the periodic slabs is $\approx 13.5$ Ang.
\label{fig:slab_geom}
}
\end{figure}

Our calculations show that DC2 in bulk is close to $E_\mathrm{F}$, which could affect the experimental AHE measurements.
An important distinction between the bulk theory and experimental transport measurements is that the experiments were carried out in a cleaved bulk terminated with an Mn surface.
Specifically, the Mn-terminated surface showed a significant modulation in the $\mathrm{d}I/\mathrm{d}V$ line map at $B = 9~T$, interpreted as a Landau quantization signature.
On the other hand, the TbSn-terminated surface showed an almost homogenous $\mathrm{d}I/\mathrm{d}V$ line map \cite{yin_quantum-limit_2020}.
This was attributed to a possible geometry reconstruction due to dangling Tb$^{3+}$ bonds \cite{yin_quantum-limit_2020}.
The Mn termination could also modify the band structure near the surface.
To explore this, we carried out DFT$+U$ calculations of a TbMn$_{12}$Sn$_{10}$ non-stoichiometric semi-infinite slab terminated with Mn layers.
In other words, the slab is periodic along in-plane directions, while a large vacuum is inserted in both out-of-plane ends, which are terminated with Mn layers, see Figure \ref{fig:slab_geom}.
To demonstrate that the Mn surface effects can be appropriately modeled by this slab, we calculated the atomic moments and charges in these two settings.
Table \ref{tab:bulk_vs_slab_mom} provides these values for Tb and Mn in various environments.
We observed that the Tb spin moments and charges are largely unchanged for bulk vs. slab, indicating that the relevant changes occur only in the surface and sub-surface Mn layers as well as Sn layers in between.
The sub-surface Mn shows a moderate change in the moments.
However, a significant change occurs near the Mn surface, where we see much larger Mn moments of $3.33~\mu_\mathrm{B}$, in quantitative agreement with previous findings \cite{lee_interplay_2022}.
This shows that the surface effects can be reliably modeled by the constructed slab.


\input{Table_2}


Figure \ref{fig:dft_opt_bands}(c) and \ref{fig:dft_opt_bands}(d) show the slab band structures.
Compared to bulk bands, the notable differences are DC3 shifting up from $\sim120$~meV to $\sim275$~meV, while DC2 shifting down very close to $E_\mathrm{F}$.
The shift in DC2 is likely due to charge transfer from the surface Mn layer to the subsurface Mn layer (see Table \ref{tab:bulk_vs_slab_mom} for atomic charges).
In Figure \ref{fig:dft_opt_bands}(c), where the SOC is not explicitly included, the DC2 is $\sim 15$~meV below the $E_\mathrm{F}$.
However, once the SOC is included, the Chern gap of $\Delta = $ 26~meV opens up, and the Fermi level lies within this gap (Figure \ref{fig:dft_opt_bands}(d)).
In reality, the Fermi level in the cleaved bulk might or might not exactly lie in the Chern gap of $34(2)$~meV, as the \textit{ab-initio} methods used here combined with a level of experimental input could result in larger systematic bias than 34~meV.
However, the change in $E_\mathrm{DC2}$ going from the bulk to the surface, $\Delta E_\mathrm{DC2} = 135$~meV must be a more robust quantity, as this is an energy difference, and we expect it to be much less sensitive to the quality of DFA or trial wave functions.


\begin{figure*}[!htbp]
\centering
\includegraphics[width=\textwidth]{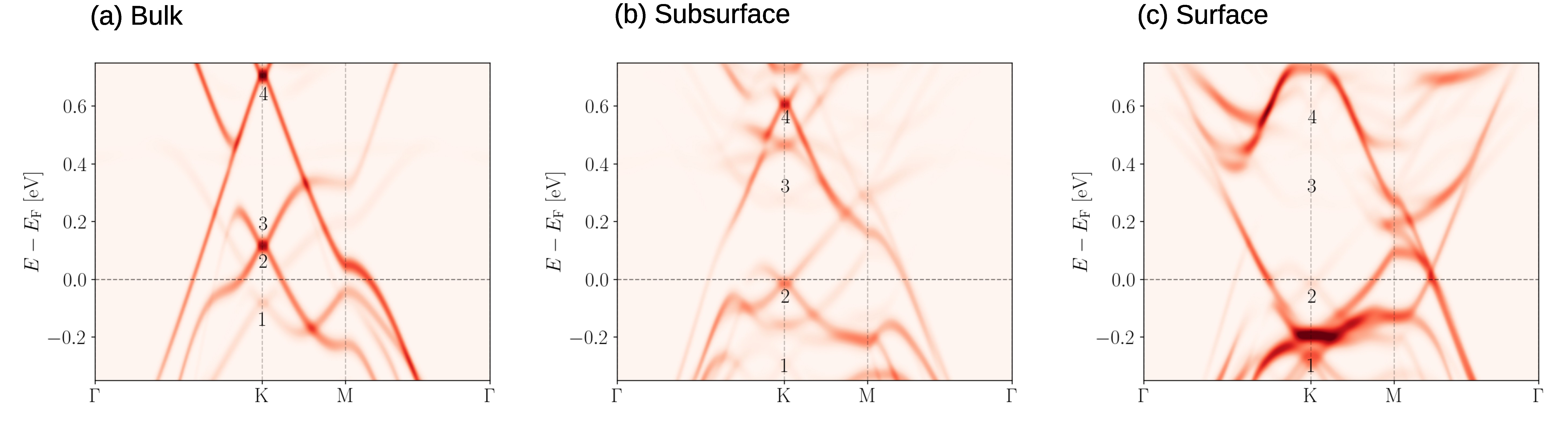}
\caption{
Minority spin band structures using projections on the atomic Mn $(d_{xy} + d_{x^2 - y^2})$ orbitals.
Projections in
(a) bulk,
(b) slab subsurface,
and
(c) slab surface
are shown.
}
\label{fig:fat_bands_compare}
\end{figure*}

To obtain further insights into the DCs' origin, we investigated their orbital characters using projections onto the atomic Mn orbitals.
In the bulk, there are two DCs near 120~meV (Figure \ref{fig:dft_opt_bands}(a), DC2 and DC3).
DC3 has a heavy $d_{zx}$ and $d_{zy}$ character, while DC2 has a $d_{xy}$ and $d_{x^2-y^2}$ character (see Supplementary Figure 9).
Since DC3 is significantly above the $E_\mathrm{F}$ in the slab ($\sim 275$~meV), we focus on DC2 with $d_{xy}$, $d_{x^2-y^2}$ character and plot the band structures of the surface, subsurface, and bulk Mn layers projected onto atomic $d_{xy}$, $d_{x^2-y^2}$ orbitals in Figure \ref{fig:fat_bands_compare}.
The subsurface Mn bands look similar to the bulk Mn bands, but the DC2 appears closer to $E_\mathrm{F}$.
On the other hand, the surface bands look quite different, and the $d_{xy}$, $d_{x^2-y^2}$ DC2 signal near the $E_\mathrm{F}$ is very weak.
We note that a DC2 near $E_\mathrm{F}$ can still be seen on the surface Mn layer, albeit with strong $d_{z^2}$, $d_{zx}$, and $d_{zy}$ characters (Supplementary Figure 11).
Namely, the DC2 near $E_\mathrm{F}$ in the slab (Figure \ref{fig:dft_opt_bands}(c)) is split between the surface and subsurface DCs in the $d$-orbital manifold with different characters. 
Perhaps it is not surprising that the surface DC2 takes on a substantial $ z$ character since it now faces the vacuum.
Assuming that STM transport measurements involved conduction mainly in $x$, $y$ directions (in-plane), the above DC2 split suggests that the experimental signal mainly consists of the bulk and subsurface layers with a stronger $d_{xy}$, $d_{x^2-y^2}$ character, and a weaker signal coming from the surface Mn $d_{zx}$, $d_{zy}$ orbitals.
This is corroborated by the fact that the STM Landau fan diagram obtained from an Mn-terminated bulk shows the Chern gap at $130(4)$~meV above $E_\mathrm{F}$, not near $E_\mathrm{F}$ as we see for the surface.

Another aspect of this material studied by Ref. \cite{lee_interplay_2022} was the $k_z$ dispersion of the bulk TbMn$_6$Sn$_6$.
Specifically, the authors mentioned that the $k_z$-dispersion was too high in the bulk TbMn$_6$Sn$_6$, which is not desired for a 2D model.
Indeed, we also find this to be true in bulk, and the band structure shows drastic changes as the $k_z$ is slowly varied (Supplementary Figure 12).
In fact, the DC2 and DC3 occur only in $k_z = 0$ when looked at increments of $\Delta k_z = 0.1$ in reciprocal lattice units and disappears for higher values.
The 'flat' band in $k_z = 0$ also gradually disperses for high $k_z$ values.
The non-ideal 2D nature of the bulk can be seen in the Fermi surface as well (see Supplementary Figure 5).
This demonstrates the additional challenge in realizing the Chern magnetism in the bulk Mn layers and motivates the experimental synthesis of TbMn$_6$Sn$_6$ thin films to eliminate the heavy $k_z$-dependance and to move the DC2 closer to $E_\mathrm{F}$. 

Finally, besides $E_\mathrm{DC2}$, we have also calculated the Dirac velocity ($v_\mathrm{D}$) to compare with experiments.
To obtain $v_\mathrm{D}$, we use the Dirac dispersion in the presence of a Chern gap due to SOC \cite{yin_quantum-limit_2020}:
\begin{equation}
    \label{eqn:dirac_disper}
    E_k = E_\mathrm{DC} \pm \sqrt{(\Delta / 2)^2 + (\hbar \mathbf{k} v_\mathrm{D})^2}.
\end{equation}
The fit of Equation \ref{eqn:dirac_disper} to the upper band of DC2 in Figure \ref{fig:dft_opt_bands}(b) provides $v_\mathrm{D} = 2.25\times 10^5$~m/s (see Supplementary Figure 4).
This is only in qualitative agreement with the experimental value of $v^\mathrm{exp}_\mathrm{D} = 4.2(3)\times 10^5$~m/s which was obtained by a fit to the Landau fan diagram \cite{yin_quantum-limit_2020}.
In addition, the Dirac velocity extracted from tunneling data provided results close to $v^\mathrm{exp}_\mathrm{D} = 4.2(3)\times 10^5$~m/s \cite{yin_quantum-limit_2020}.
Therefore, LDA$+U^{3d}_\mathrm{Mn}$ seems to genuinely underestimate $v_\mathrm{D}$ (and relatedly, Fermi velocities).
This $v_\mathrm{D}$ underestimation is only mildly sensitive to the value of $+U^{3d}_\mathrm{Mn}$, as the increase of $v_\mathrm{D}$ going from LDA$+U^{3d}_\mathrm{Mn}$($-0.5$ eV) to LDA$+U^{3d}_\mathrm{Mn}$($+0.5$ eV) is only $\Delta v_\mathrm{D} = 0.27\times 10^5$~m/s.
Namely, the main effect of $+U^{3d}_\mathrm{Mn}$ in this material is to shift the band energies up/down while only moderately changing the dispersions (namely slopes or $v_\mathrm{D}$).
Therefore, we don't expect LDA$+U^{3d}_\mathrm{Mn}$($-0.5$ eV) to obtain both $E_\mathrm{DC2}$ \textit{and} $v_\mathrm{D}$ correctly, nor do we expect it to produce accurate results for all other properties such as optimized geometries.
We note that the $v_\mathrm{D}$ underestimation is not only specific to TbMn$_6$Sn$_6$, and it has been observed in other materials such as graphene and Dirac semimetal Na$_3$Bi, where LDA underestimates the Fermi velocity while hybrid DFT or $GW$ methods show considerable improvements \cite{trevisanutto_ab_2008, di_bernardo_importance_2020}.
Ultimately, the inability to predict all properties using the same DFA shows the limitations of the single-particle picture, which supports our findings throughout this work.

To further investigate the nature of electron correlations and spin fluctuations in TbMn$_6$Sn$_6$, we carried out DFT+Dynamical Mean Field Theory (DMFT) calculations (see Supplementary Note 5 for DMFT methods).
The preliminary results on the high-temperature paramagnetic (PM) phase of TbMn$_6$Sn$_6$ show strong orbital-dependent electron correlations.
We find that in the PM phase, the electron occupation of the Mn $3d$-subspace to be $5.3~e^-$ and the mean fluctuating local moments ($<m_z>$) of Mn $3d$ electrons to be $\sim 3.7~\mu_\mathrm{B}$, which is in good agreement with a recent DFT+DMFT computation performed on YMn$_6$Sn$_6$ \cite{li_dirac_2021}, a sister compound in this family.
To compute the fluctuating moment, we use
\begin{equation}
<m_z> = 2\sum_i P_i|S_z|_i
\end{equation}
where $P_i$ is the probability of the $i^\mathrm{th}$ multiplet in the continuous time Monte Carlo impurity solver and $|S_z|_i$ is the absolute value of the corresponding moment.
The histogram plot (Supplementary Figure 16) shows that the most probable spin states are high-spin states with $S_z=2.0$, $2.5$, and $1.5$, indicating the correlation in the PM phase is likely due to Hund’s rule coupling \cite{medici_hunds_2017}.
This prediction also agrees with other kagome metals, including YMn$_6$Sn$_6$ \cite{li_dirac_2021, huang_signatures_2020}.
Hund's metals were shown to display strong electron correlations \cite{georges_strong_2013}, and this agrees with our QMC results, which show significant correlations due to multi-reference character.
The underlying mechanism of the origin of ferromagnetism with the ordering of fluctuating moments is an important area for further studies and is beyond the scope of the present work.

\section{Discussion}
\label{sec:discus}

This study has made advancements in three distinct but related directions.
First, an accurate, correlation-consistent ECP was generated for the Tb element with a valence space of $4f^9 5s^2 5p^6 6s^2$.
To the best of our knowledge, the accuracy of such a core-valence partitioning was not known for many-body methods such as CCSD(T) or DMC. 
In CCSD(T), we found that the low-lying atomic gaps and the binding energies of TbH$_3$ and TbO can be made chemically accurate (1 kcal/mol) with careful optimizations. 
In light of the obtained results, this seemingly too technical aspect of the effective core model is actually important for a clear delineation of subtle physical effects in the studied system. 
Single-reference DMC calculations showed that $\approx 12\%$ of the correlation energy is missing in Tb, TbH$_3$, and TbO.
We believe the promising results obtained for the Tb element indicate that similarly accurate ccECPs could be generated for other rare-earth elements.
Such a set of rare-earth ccECPs would open further possibilities to study the broader family of $R$Mn$_6$Sn$_6$ with significantly improved accuracy using many-body methods such as CCSD(T), CI, QMC, and also DFT with appropriate DFAs.
In fact, the current progress in applying real-space QMC to $f$-element systems seems to be hindered by the unavailability of accurate enough rare-earth ECPs \cite{bauschlicher_reliability_2022, hegde_quantifying_2022}, as we were able to find only a few QMC studies in the literature \cite{elkahwagy_diffusion_2016, elkahwagy_diffusion_2017, elkahwagy_theoretical_2018, devaux_electronic_2015}.
Therefore, we hope this work will motivate further studies in this avenue.

Second, recent studies \cite{fu_applicability_2018, fu_density_2019} showed that meta-GGA such as SCAN, hybrid DFT such as PBE0, and DFT$+U$ methods overestimate the magnetic moments in simple elemental metals such as bcc Fe, hcp Co, fcc Ni, and other transition metals.
Here, we observe a similar overestimation in Mn magnetic moments when using DFT$+U$.
Importantly, we show that single-reference QMC calculations also severely overestimate the transition metal moments similar to the previously shown \cite{fu_applicability_2018, fu_density_2019} advanced DFAs such as SCAN and PBE0.
The overestimation persists even when the dynamic correlations are extrapolated to the zero-variance limit, suggesting that a multi-reference treatment is needed to predict the magnetic moments correctly.
These results have broader implications for the ability of commonly used DFAs, such as LDA, to effectively capture the multi-reference effects and for the origins of effective weak interactions in metallic systems.
A study of static correlation effects is possible for such systems with small primitive cells as multi-reference methods such as CCSD(T) and CI mature for use in periodic boundary conditions.
We leave this aspect of the study to future work and hope that these findings will stimulate further research in this direction.

Finally, using a combination of \textit{ab-initio} and neutron diffraction experimental results, we reveal key insights about TbMn$_6$Sn$_6$ bulk and non-stoichiometric thin film limit.
We show that DC2 is only $\sim 120$ meV above $E_\mathrm{F}$ in bulk, in agreement with experiments \cite{yin_quantum-limit_2020}.
However, the realization of Chern magnetism in bulk is complicated by the heavy $k_z$-dispersion as suggested previously \cite{lee_interplay_2022, jones_origin_2022}.
We show that in the slab, the DC2 shifts even closer to $E_\mathrm{F}$.
This motivates further experimental studies of TbMn$_6$Sn$_6$ in the thin film limit to eliminate the $k_z$-dispersion and to probe for the possible realization of Chern magnetism.
Viable pathways to experimentally synthesize defect-free \cite{zunger_beware_2019} thin films in a controlled manner to realize the Chern phase remains to be shown.
Our work thus illuminates the aspects of TbMn$_6$Sn$_6$ both from theoretical and experimental viewpoints and opens the door for future studies of this exciting material.


\section{Methods}
\label{sec:methods}

\subsection{DFT$+U$ Methods}
DFT+$U$ calculations were carried using \textsc{quantum espresso} package \cite{giannozzi_quantum_2009, giannozzi_advanced_2017, giannozzi_quantum_2020}.
DFT+$U$ is well-known to converge to meta-stable states depending on the given initial guess for the Hubbard occupation matrix.
To overcome this issue, we used a method similar to the ramping method described in Ref. \cite{meredig_method_2010}.
Specifically, first, we converged a DFT$+U$ calculation with a very small $U$ value, such as $10^{-16}$ eV.
Then, the converged charge density, orbitals, and Hubbard occupation matrices were directly provided as an initial guess for the desired value of $U$, skipping the adiabatic ramping of $U$.
On a few occasions, we tested that this method indeed results in lower energies than the default values for Hubbard occupations.

\subsection{QMC Methods}
QMC calculations were carried out using the \textsc{qmcpack} package \cite{kim_qmcpack_2018, kent_qmcpack_2020}.
Most calculations were driven by the \textsc{nexus} automation tool \cite{krogel_nexus_2016}.
All bulk QMC calculations use canonical twist-averaging (CTA) with $[8 \times 4 \times 8]$ twists to sample the kinetic energy.
In this approach, every twist is charge-neutral.
In addition, every twist ($\theta$) is enforced to have the same cell magnetization:
\begin{equation}
M^\mathrm{s}(\theta) = N_{e}^{\uparrow}(\theta) - N_{e}^{\downarrow}(\theta) = \mathrm{Constant}
\end{equation}
namely, every twist has the same magnetization as the twist-averaged magnetization of the cell.
At each twist, the orbitals are occupied using the lowest Kohn-Sham eigenvalues in each spin channel.
All molecular and bulk Jastrow functions were optimized using energy minimization in VMC.
For bulk calculations, the Jastrow was optimized at $k = \Gamma$ twist and reused for other twists.
This optimization was carried out for LDA orbitals at $M^\mathrm{s} = 7~\mu_\mathrm{B}$ and reused for all other DFAs and $M^\mathrm{s}$ values.
This was done to obtain improved energy differences originating from the determinantal part of the trial wave function.

The UCCSD(T) calculations were carried out using the \textsc{Molpro} package \cite{werner_molpro_2012}.
For COSCI calculations, we used the \textsc{dirac} code \cite{saue_dirac_2020}.

\section{Data Availability}
See the Supplementary Information for extended calculations details, such as 
employed bulk and slab geometries (Supplementary Note 1), 
pseudopotential optimization (Supplementary Note 2), 
DFT$+U$ methods and data (Supplementary Note 3),
QMC methods and data (Supplementary Note 4),
and
DMFT methods and data (Supplementary Note 5).
The Supplementary Information includes additional References \cite{hay_ab_1998, stoll_relativistic_2002, bergner_ab_1993, igel-mann_pseudopotentials_1988, burkatzki_energy-consistent_2007, trail_norm-conserving_2005, lajohn_ab_1987, stevens_relativistic_1992, wang_new_2019, annaberdiyev_new_2018, casula_beyond_2006, mitas_nonlocal_1991, haule_dynamical_2010, blaha_wien2k_2019, ExactDC, metal-SM}.

The input files, output files, and supporting data generated in this work are published in Materials Data Facility~\cite{blaiszik_materials_2016, blaiszik_data_2019} and can be found in Ref.~\cite{mdf_data}.
The data is also available from the corresponding author upon reasonable request.

\begin{acknowledgments}

We thank P. R. C. Kent, Anand Bhattacharya, and Jeonghwan Ahn for reading the manuscript and providing helpful suggestions.

This work has been supported by the U.S. Department of Energy, Office of Science, Basic Energy Sciences, Materials Sciences and Engineering Division, as part of the Computational Materials Sciences Program and Center for Predictive Simulation of Functional Materials.


An award of computer time was provided by the Innovative and Novel Computational Impact on Theory and Experiment (INCITE) program.
This research used resources of the Oak Ridge Leadership Computing Facility, which is a DOE Office of Science User Facility supported under Contract No. DE-AC05-00OR22725.
This research used resources of the National Energy Research Scientific Computing Center (NERSC), a U.S. Department of Energy Office of Science User Facility located at Lawrence Berkeley National Laboratory, operated under Contract No. DE-AC02-05CH11231. S. M. acknowledges the support from the Air Force Office of Scientific Research by the Department of Defense under the award number FA9550-23-1-0498 of the DEPSCoR program and the Frontera supercomputer at the Texas Advanced Computing Center (TACC) at The University of Texas at Austin, which is supported by National Science Foundation grant number OAC-1818253.

This paper describes objective technical results and analysis. Any subjective views or opinions that might be expressed in the paper do not necessarily represent the views of the U.S. Department of Energy or the United States Government.

Notice:  This manuscript has been authored by UT-Battelle, LLC, under contract DE-AC05-00OR22725 with the US Department of Energy (DOE). The US government retains and the publisher, by accepting the article for publication, acknowledges that the US government retains a nonexclusive, paid-up, irrevocable, worldwide license to publish or reproduce the published form of this manuscript, or allow others to do so, for US government purposes. DOE will provide public access to these results of federally sponsored research in accordance with the DOE Public Access Plan (http://energy.gov/downloads/doe-public-access-plan).

\end{acknowledgments}

\section{Competing Interests}
The authors declare no competing financial or non-financial interests.

\section{Author Contributions}
P.G. conceived and supervised the project.
A.A. carried out the majority of calculations and wrote the manuscript.
S.M. carried out the DFT+eDMFT calculations and wrote the related text.
L.M. and J.T.K. aided with methodology approaches and writing the manuscript.

\bibliographystyle{naturemag}
\bibliography{main.bib}

\end{document}

%% file: Table_1.tex
\begin{table}[!htbp]
\centering
\caption{
Mn moments and cell magnetizations $M^\mathrm{s}$ for various DFAs.
The fraction of exact exchange (EXX) in the DFAs is shown.
Note that as EXX is increased, the magnetizations are increasingly overestimated.
}
\label{tab:hybrids_mag}  
\begin{tabular}{lccc}
\hline
DFA   & EXX  & Mn $[\mu_\mathrm{B}]$ & $M^\mathrm{s}$ $[\mu_\mathrm{B}]$ \\
\hline
LDA   & 0.00 & 2.51 & 7.09  \\
PBE   & 0.00 & 2.76 & 8.17  \\
B3LYP & 0.20 & 3.38 & 10.81 \\
PBE0  & 0.25 & 3.47 & 11.00 \\
HF    & 1.00 & 4.34 & 14.09 \\
\hline
Exp. \cite{el_idrissi_magnetic_1991, jones_origin_2022} &
             & 2.39(8) & 6.86 \\
\hline
\end{tabular}
\end{table}

%% file: Table_2.tex
\begin{table}[!htbp]
\centering
\caption{
Comparison of atomic moments and charges in bulk vs. slab.
Scalar-relativistic LDA$+U^{3d}_\mathrm{Mn}$($-0.5$ eV) results.
Note that Tb quantities are very close, signifying that slab calculation is a good model for the Mn-surface cleaved bulk.
}
\label{tab:bulk_vs_slab_mom}
\begin{tabular}{llcc}
\hline
Atom & Environment     & Moment [$\mu_\mathrm{B}$] & Charge [$e^{-}$] \\
\hline
Tb   & Slab            & -6.407 & 1.173 \\
Tb   & Bulk            & -6.403 & 1.175 \\
Mn   & Slab surface    &  3.333 & 1.476 \\
Mn   & Slab subsurface &  2.315 & 1.431 \\
Mn   & Bulk            &  2.366 & 1.437 \\
\hline
\end{tabular}
\end{table}